# Predicting brain tumour enhancement from non-contrast MR imaging with artificial intelligence: a multi-cohort retrospective diagnostic accuracy study

James K. Ruffle FRCR PhD*, Samia Mohinta MSc, Guilherme Pombo PhD, Asthik Biswas FRCR†, Alan Campbell FRCR†, Indran Davagnanam FRCR†, David Doig FRCR†, Ahmed Hammam FRCR†, Harpreet Hyare FRCR PhD†, Farrah Jabeen FRCR†, Emma Lim FRCR†, Dermot Mallon FRCR PhD†, Stephanie Owen FRCR†, Sophie Wilkinson FRCR†, Sebastian Brandner FRCPath PhD, Parashkev Nachev FRCP PhD

Queen Square Institute of Neurology, University College London, London, UK (*J K Ruffle FRCR PhD, S Mohinta MSc, I Davagnanam FRCR, H Hyare FRCR PhD, D Mallon FRCR PhD, Prof S Brandner FRCPath PhD, Prof P Nachev FRCP PhD*); Lysholm Department of Neuroradiology, National Hospital for Neurology and Neurosurgery, London, UK (*J K Ruffle FRCR PhD, I Davagnanam FRCR, D Doig FRCR, A Hammam FRCR, H Hyare FRCR PhD, F Jabeen FRCR, D Mallon FRCR PhD, S Owen FRCR*); NVIDIA, UK (*G Pombo PhD*); Great Ormond Street Hospital for Children, London, UK (*A Biswas FRCR*); Imaging Department, Royal National Orthopaedic Hospital, Stanmore, Middlesex, UK (*A Campbell FRCR*); Department of Radiology, University College Hospitals NHS Foundation Trust, London, UK (*H Hyare FRCR PhD, S Wilkinson FRCR*); Department of Radiology, Royal Free Hospital, London, UK (*F Jabeen FRCR*); Department of Imaging, Imperial College Healthcare NHS Trust, London, UK (*E Lim FRCR*); Department of Brain Sciences, Imperial College London, London, UK (*E Lim FRCR*); and Division of Neuropathology and Department of Neurodegenerative Disease, Queen Square Institute of Neurology, University College London, London, UK (Prof S Brandner FRCPath PhD).

*Correspondence to:
Dr James K Ruffle
Email: j.ruffle@ucl.ac.uk
Address: Institute of Neurology, UCL, London WC1N 3BG, UK

†Contributed equally.

## Declaration of interests

GP is affiliated with NVIDIA but contributed to this work as an academic collaborator and former core member of the research team, not in any NVIDIA capacity; NVIDIA itself had no involvement in this research — including study design, data collection, analysis, interpretation, and the decision to publish — and provided no hardware or in-kind support. PN is affiliated with Hologen, a healthcare AI company, which had no involvement in this research project. All other authors declare no competing interests.

## Manuscript Type

Original article

## Authorship

Conceptualisation: JKR, PN. Methodology: JKR, PN. Software: JKR. Formal analysis: JKR, SM, GP, PN. Investigation: JKR, SM, GP, SB, AB, AC, ID, DD, AH, HH, FJ, EL, DM, SO, SW. Data curation: JKR, SM, GP, SB. Writing – original draft: JKR, PN. Writing – review & editing: all authors. Visualisation: JKR. Supervision: PN. Project administration: JKR. Funding acquisition: JKR, PN. Both JKR and PN have accessed and verified the data and were responsible for the decision to submit the article.

## Abbreviations

AI, artificial intelligence; BA, balanced accuracy; FLAIR, fluid attenuated inversion recovery; MRI, magnetic resonance imaging; NHNN, National Hospital for Neurology and Neurosurgery; SD, standard deviation.

## Abstract

**Background:** Brain tumour imaging assessment typically requires both pre- and post-contrast MRI, but gadolinium administration is not always desirable, such as in frequent follow-up, renal impairment, allergy, or paediatric patients. We aimed to develop and validate a deep learning model to predict brain tumour contrast enhancement from non-contrast MRI sequences alone.

**Methods:** We assembled 11,089 brain MRI studies (acquired 2006–2024) from 10 international datasets across four countries and three continents (North America, Europe, and Africa), spanning adult and paediatric populations (no age limit) with glioma, meningioma, metastases, and post-resection appearances. Three deep-learning architectures were trained to detect and segment enhancing tumour from T1w, T2w and T2/FLAIR sequences alone. Performance was assessed in a 1,109 held-out test set (primary endpoint: patient-level enhancement detection; secondary endpoint: voxel-level Dice). Eleven expert radiologists attempted the same novel task on a 564-case subset, reviewing 100 cases each, and were blinded to clinical history, prior imaging, and referral indication.

**Findings:** The best performing model**,** nnU-Net, achieved 83.0% balanced accuracy (95% CI 79.1–87.2; sensitivity 91.5% [89.9–93.1], specificity 74.4% [66.7–82.5]) for enhancement detection, with $R^2 = 0.859$ for enhancement-volume prediction. 762/992 (76.8%) of enhancing cases reached Dice ≥ 0.3, 670/992 (67.5%) Dice ≥ 0.5, and 498/992 (50.2%) Dice ≥ 0.7. In 564 matched cases, the radiologists' majority-vote performance was lower, with a balanced accuracy of 71.7% (sensitivity 77.6%, specificity 65.8%) under the same blinded conditions. Patient-level enhancement detection was generally performant across pathologies (proportion of patients reaching voxel-level Dice ≥0.3: meningioma 93 of 100 [93%], presurgical glioma 526 of 691 [76%], metastases 52 of 70 [74%], postoperative glioma 82 of 111 [74%]) but was lower for paediatric cases (9 of 20 [45%]).

**Interpretation:** Deep learning can identify contrast-enhancing brain tumours from non-contrast MRI. These models show promise as a triage or decision-support adjunct — for example, flagging studies in which post-contrast enhancement is likely so that contrast can be added to a non-contrast-only protocol — and may reduce gadolinium dependence in neuro-oncology imaging. Future work should optimise such models in collaboration with radiology experts.

**Funding:** European Society of Radiology; European Institute for Biomedical Imaging Research (EIBIR); Medical Research Council (MR/X00046X/1 & UKRI1389); Wellcome Trust (213038/Z/18/Z); UCLH NIHR Biomedical Research Centre.

## Research in context

**Evidence before this study:** We searched PubMed, Embase, and Web of Science from database inception to 1 April 2026 using combinations of (“brain MRI” OR “brain tumour” OR “glioma”) AND (“deep learning” OR “convolutional neural network” OR “U-Net” OR “transformer”) AND (“contrast enhancement” OR “gadolinium” OR “virtual contrast” OR “synthetic contrast”). No language restrictions were applied; the reference lists of retrieved studies were also screened. Previous research had demonstrated the feasibility of predicting or synthesising contrast enhancement from non-contrast brain MRI using deep learning, but the available evidence was confined to single-institution, adult, predominantly glioma cohorts of modest size, with most studies enrolling tens to hundreds of patients and very few conducted at larger scales. Reported performance metrics are heterogeneous across studies, spanning synthetic-image quality indices (structural similarity index and peak signal-to-noise ratio) and enhancing-tumour Dice coefficients, with no consistent patient-level detection benchmarks. Other earlier works on gadolinium dose reduction have shown that deep learning could preserve image quality at approximately four- to ten-times reduced contrast-agent dose, though such work still requires gadolinium administration.

**Added value of this study:** To our knowledge, this is the largest and most heterogeneous evaluation of contrast-enhancement prediction from non-contrast brain MRI to date, and the first to directly benchmark deep learning against a blinded expert-radiologist panel on the same non-contrast task, to include a paediatric cohort, and to report performance stratified by demographic and geographic subgroup with a formal equity assessment. We additionally quantify the drivers of detection failure (of 260 failing cases: 105 small lesions $<0.5$ cm$^3$ [40.4%], 100 atypical enhancement [38.5%], 30 false positives [11.5%], 17 postoperative change [6.5%], 8 paediatric cases [3.1%]) and provide a descriptive characterisation of predictive entropy. The approach proved feasible at scale across ten international cohorts and five pathology classes, while remaining weakest in the paediatric population — establishing both its promise and its current limits.

**Implications of all the available evidence:** Deep learning prediction of contrast enhancement from non-contrast brain MRI achieved higher balanced accuracy than expert radiologists on this unfamiliar non-contrast task. Although performance remains currently insufficient to replace contrast-enhanced MRI acquisitions outright, particularly in the paediatric population, where the model was weaker, it indicates a direction of travel in which pertinent imaging findings previously only delineable from post-contrast imaging can be derived from an AI trained to elicit them from non-contrast imaging alone. Used as a triage or decision-support adjunct, such a model could flag studies in which post-contrast enhancement is likely, prompting addition of contrast to a non-contrast-only study, thereby avoiding a second hospital visit and minimising delays to patient care, while imaging predicted to be enhancement-free might, with appropriate caution, proceed without gadolinium. Taken together with prior single-institution literature, these findings are consistent with academic and clinical consensus on measures to reduce gadolinium dependence in imaging assessments, including for oncology.

# Introduction

Imaging diagnosis and surveillance play a key role in neuro-oncological care across both adult and paediatric populations[1]. The presence or absence of contrast enhancement is often critical to clinical decision-making. Contrast agents, however, cannot always be given, owing to contraindications such as allergies or renal impairment, or patient refusal. Imaging may also be corrupted by artefacts, making its radiological interpretation difficult. Furthermore, it is often desirable to minimise repeated gadolinium use, especially in paediatric patients, though the risk of inadequate tumour characterisation is typically perceived to outweigh the risk of repeated gadolinium administration. Known risks aside, such as the development of nephrogenic systemic fibrosis, the long-term effects of gadolinium exposure and its deposition in the body remain unknown[2].

Beyond the individual patient, the widespread use of gadolinium contrast agents in medical imaging has a broader environmental impact. Although only small amounts are used in each study, the cumulative impact at a global scale results in substantial amounts (in the order of many thousands of litres) of gadolinium administered, secreted and/or disposed of each year[3]. Gadolinium has been detected in sewage, surface, drinking water, water sources distant from MRI facilities, and even in fast-food soft drinks. Given uncertainty about the long-term impact of exposure, even at low doses, the problem has stimulated worldwide research efforts to reduce, or even eliminate, its use.

One possible solution to reducing contrast dependence is to use artificial intelligence to predict the presence of enhancement from non-contrast imaging alone. We have recently demonstrated a proof of concept that deep learning models can segment enhancing tumour tissue in glioma without contrast-enhanced imaging[4]. Similar approaches have also been investigated by others, both within the brain and in other body parts, to either reduce gadolinium dosage[5,6] or eliminate it entirely for lesion detection or image reconstruction[6,7]. Taken together, there is strong support that complex models can leverage non-contrast MRI sequence information to derive post-contrast information from non-contrast scans.

Here, we prototype an enhanced intracranial disease prediction system using only non-contrast MRI data from the largest known neuro-oncology imaging cohort, combining state-of-the-art deep learning with large-scale, unselective neuro-oncology data. Beyond scale, our methodological contributions include: (1) a systematic comparison of convolutional neural network and transformer-based architectures; (2) descriptive characterisation of predictive entropy; (3) comprehensive equity assessment across demographic, clinical, and country-of-origin subgroups; and (4) direct benchmarking against expert human radiologist performance. Our findings strongly support the notion that complex models can learn features of what has historically been referred to as ‘enhancing’ disease when only non-contrast imaging is provided.

# Methods

## Data

We combined in-house and international open source brain tumour data to study a final dataset of 11,089 imaging sessions from 8,548 unique patients diagnosed with a brain tumour (Figure 1; Appendix p1). This cohort was as maximally heterogeneous and inclusive as possible, spanning adult and paediatric patients, pre- and post-operative imaging, multiple neuro-oncological pathologies, and numerous global study sites (Appendix p1, p6, p8). Imaging was acquired at multiple field strengths (1.5T and 3T) and from multiple MRI manufacturers between 2006 and 2024. Eligibility was deliberately broad: any patient with a brain-tumour MRI study available in the contributing datasets was eligible, with no age limit (spanning paediatric to adult patients) and no restriction by histology, molecular profile, performance status, or comorbidity. As this was a retrospective analysis of existing imaging, trial-specific eligibility criteria—formal disease-response evaluation, performance status, estimated life expectancy, eligibility laboratory tests, and prior-treatment or washout requirements—were not applicable. No patients were excluded on eligibility grounds. The cohort comprised all available brain-tumour MRI studies. Cases failing image-quality control (inter-sequence motion or registration failure) or flagged as cross-dataset duplicates ($r > 0.95$) were removed at preprocessing. The public cohorts were pre-curated by their providers, and per-dataset screening counts were not separately re-logged in this secondary analysis. Patients and members of the public were involved in discussing the research question and its study design, but not the choice of outcome measures, the conduct of the study, the analysis or interpretation of the data, or the writing of the report.

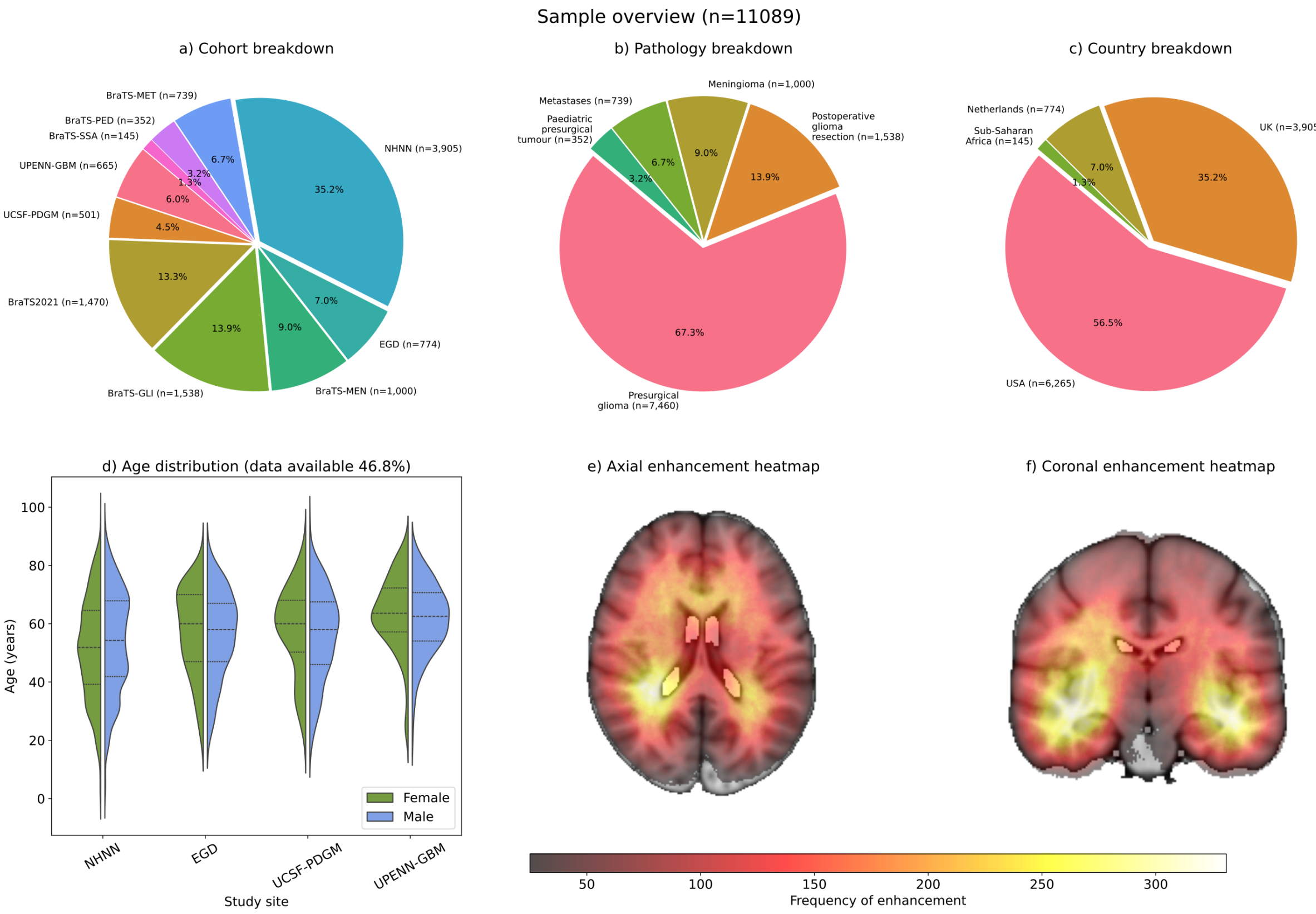

**Figure 1. Sample overview.** a) Cohort breakdown across the ten different study sites. b) Pathology type breakdown. c) Country of sample origin breakdown. d) Cohort age and sex distribution, where this data was available. e-f) Axial and coronal enhancing lesion distribution map with warmer colours indicating higher frequency of enhancement occurrence within a given location. Abbreviations: NHNN, National Hospital for Neurology and Neurosurgery; UPENN-GBM, University of Pennsylvania Glioblastoma Dataset; UCSF-PDGM, University of California San Francisco Preoperative Diffuse Glioma Dataset; EGD, Erasmus Glioma Dataset; BraTS, Brain Tumour Segmentation Challenge; BraTS-GLI, Post-treatment Glioma; BraTS-MEN, Meningioma; BraTS-MET, Metastases; BraTS-PED, Paediatric; BraTS-SSA, Sub-Saharan Africa.

## Model development

We trained a series of fully volumetric multi-modal 3D deep learning models to segment enhancing tumours from non-contrast imaging sequences (Figure 2). Models were provided with the ground-truth labels of background (i.e., non-brain), normal brain parenchyma, non-enhancing abnormal tissue, and enhancing neuro-oncological disease, the latter being the primary target of interest. The derivation of the ground truths is further described within Appendix p2, but notably, the enhanced tumour component was always derived from complete datasets, i.e., including post-contrast T1-weighted sequences.

Here, however, the model input consisted only of non-contrast sequences, specifically T1w-, T2w-, and T2/FLAIR-weighted MRI sequences. We performed a randomised train-test split at the patient level, with 90% of patients allocated to training and 10% to model testing, ensuring no leakage between sets. The test set was locked until all model development was complete. Site and scanner identifiers were not supplied to the model as inputs, and no site- or scanner-level intensity harmonisation was applied, so that the model learned from rather than being shielded from acquisition heterogeneity (Appendix p2). Trained model architectures included nnU-Net (v2)[8], SegResNet[9], and SwinUNETR[10] (Appendix p3-5).

## Evaluation

### Performance metrics

We evaluated the performance of trained segmentation models on a held-out test set. Model performance was assessed using both voxel-level and patient-level metrics. Voxel-wise metrics, computed in 3D across the full volumetric prediction and ground truth for each imaging session, included the Dice coefficient, balanced accuracy, precision, recall, and F1 score. Bootstrapped (1000 iterations) patient-wise metrics were also derived for the overall classification of whether a study contained any pathologically enhancing tissue, quantified by balanced accuracy, macro-averaged precision, recall, and F1 score. At the patient level, a case was classified as 'enhancement detected' if the model correctly predicted any enhancing voxels after thresholding the softmax probability map at 0.5. For clinical interpretability of segmentation quality, we further refined patient and voxel-level detection success thresholds as follows: Dice ≥ 0.3 ('acceptable detection'); Dice ≥ 0.5 ('good detection'); and Dice ≥ 0.7 ('excellent detection'). Below Dice 0.3, spatial predictions were considered clinically unreliable and served as a heuristic to ensure that models were not overly sensitive and did not exhibit high false-positive rates for enhancement detection. Our rationale for including both voxel-wise metrics and patient-level enhancement detection metrics was to encompass both the conventional statistical evaluations undertaken in computer vision research (which typically operate at the voxel level) and those in clinical research (which typically operate at the patient level).

### Equity assessment

We quantified the epistemic equity of the model[11], i.e., fairness in performance across patient subgroups, ensuring equitable diagnostic utility regardless of patient characteristics. We assessed variation in fidelity across patient attributes as a function of its performance, considering the cohort site, the underlying pathology, the country of sample origin, the age and sex of the patient (where available)[12], and the size and morphology (Appendix p4-5). In line with the Sex and Gender Equity in Research (SAGER) guidelines, sex was ascertained directly from the source datasets and denotes biological sex rather than self-reported gender. Race and ethnicity were not available in any of the source datasets, so they could not be evaluated as equity attributes.

### Comparison with radiologists

We subsequently conducted an assessment in which expert radiologists were tasked with predicting whether a lesion would enhance after contrast administration, using only the unenhanced images. To do this, 100 patients from the held-out test set were randomly allocated to each of the eleven individual experienced radiologists. The eleven radiologists comprised board-certified specialists (post-FRCR, post-CCT) with either formal neuroradiology subspecialty training or extensive neuro-oncological imaging experience (median years in practice: 11.3, IQR 7.5–14.5). All reported brain tumour MRI as part of routine clinical practice across six large London-based hospitals. Cases were randomly sampled from the held-out test set, with each radiologist reviewing exactly 100 cases (50 with an enhancing abnormality and 50 without), yielding 1,100 total case-reader evaluations. Randomly drawn, this yielded 564 unique imaging sessions from the 1,109-study test set, of which 447 contained an enhancing tumour, and 117 did not. Some sessions were reviewed by multiple readers, others by one. Radiologists accessed anonymised imaging through ITKSnap on a Linux workstation with a 4K monitor. No time limit was imposed. Readers were informed that approximately half of the cases would contain enhancing lesions for statistical purposes, but were not provided with additional clinical context. Cases were presented in a randomised order to minimise recall bias. Each radiologist sequentially reviewed the unenhanced MRI images (T1w-, T2w-, and T2/FLAIR-weighted) from each case and was asked: “Do you think there will be an enhancing abnormality in this case?” We then statistically compared the model's predictions with those of the radiologist, with both sets related to the ground truth.

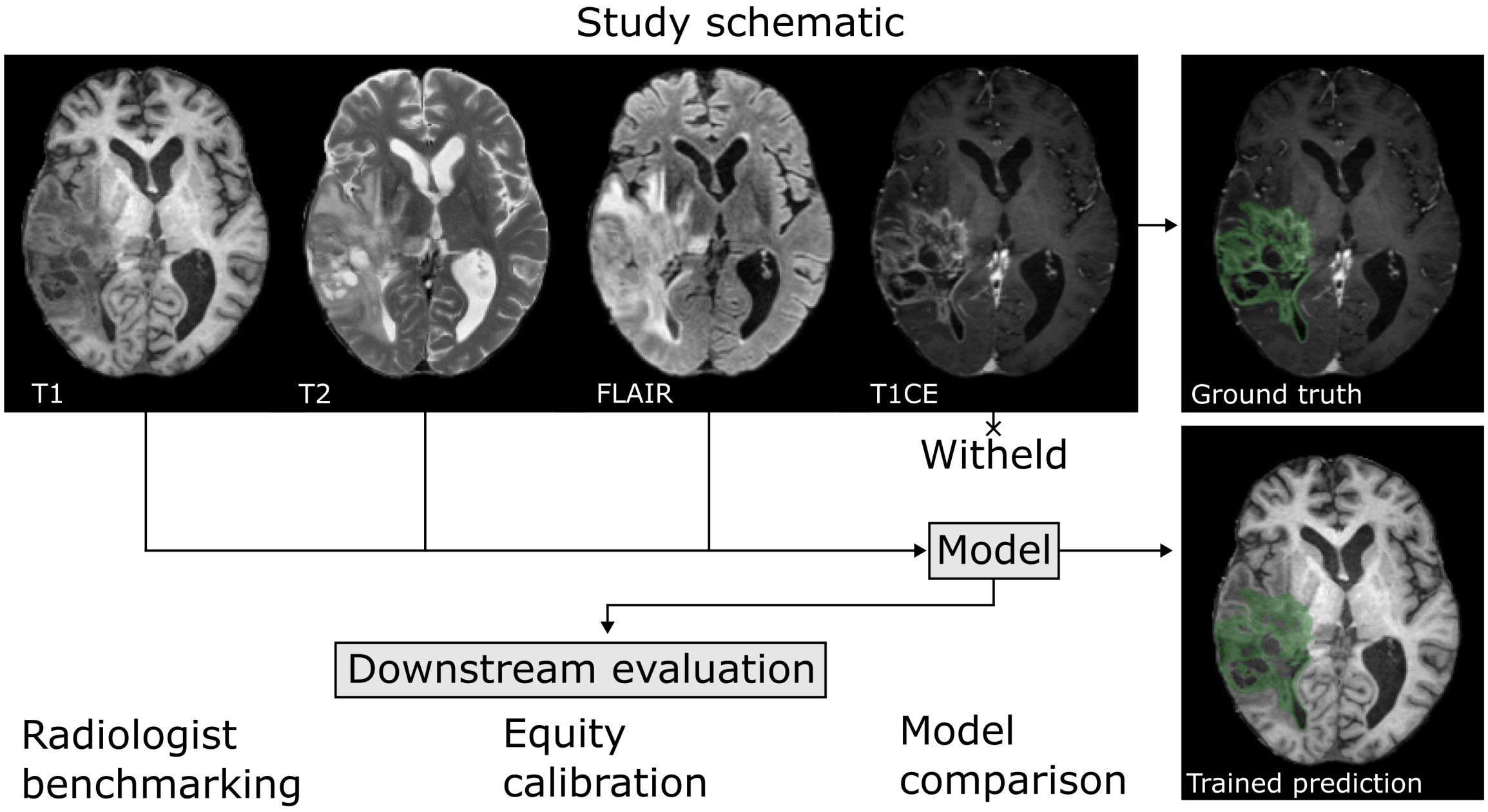

**Figure 2. Study schematic**. The model was trained to predict enhancing intracranial disease from T1w (T1-weighted), T2w (T2-weighted), and T2/FLAIR-weighted imaging alone. Ground truth labels were generated from the complete imaging datasets, including all available sequences (T1w, T2w, T2/FLAIR, and post-contrast T1

[T1CE]), with T1CE being the key sequence for enhancement delineation. The model input consisted solely of non-contrast sequences. Downstream model evaluation was undertaken by conventional model comparison, radiologist benchmarking, and equity assessments.

## Analytic compliance

All analyses were performed and reported following TRIPOD-AI and CLAIM guidelines. Classical statistical tests included logistic regression, one-way ANOVA, t-tests, Levene's test for equality of variances, and McNemar's exact test. Patient-level performance metrics (balanced accuracy, sensitivity, specificity, precision, F1) and their 95% confidence intervals were computed by patient-wise bootstrap resampling (1,000 resamples). The random seed was fixed for all analyses. Voxel-level Dice was compared across architectures using the paired Wilcoxon signed-rank test on per-case Dice, and patient-level detection metrics were compared using a paired case-level bootstrap of the between-architecture difference (10,000 resamples). The model was compared against the expert-radiologist majority vote on the matched cases using McNemar's exact test. Negative and positive predictive values and clinical utility estimates (contrast administrations potentially avoided and enhancing tumours potentially missed per 1,000 scans) were derived from sensitivity and specificity using Bayes' theorem across a range of disease prevalences. Inter-rater agreement for the multi-rater subset of those cases was summarised using Cohen's kappa across all reader-pair combinations. Statistical tests were corrected for multiple comparisons using Bonferroni adjustment, with families of contrasts tested and corrected p-values reported accordingly. All p-values are two-sided and reported as exact values where $p \geq 0.0001$ and as $p < 0.0001$ below this floor. The analysis code was cross-checked for correctness with assistance from Claude Code (Anthropic).

## Ethics

The local ethics committee at University College London approved the study. We received ethical permission for the consentless analysis of local irrevocably anonymised data collected during routine clinical care (approval reference 07/H0707/152). All external datasets were also granted ethical approval, with the consent process described within the source datasets[13-21].

## Role of the funding source

The funders of the study had no role in study design, data collection, data analysis, data interpretation, or writing of the report. The corresponding author had full access to all study data and final responsibility for the decision to submit for publication.

## Results

The study cohort included 11,089 imaging sessions, with 9980 allocated to the training dataset and 1109 assigned to model testing. Patient age was available for 5187 patients (46.78%), with a mean ± standard deviation of 55 ± 16.44 years. Patient sex was available for 5651 patients (50.96%), 3061 of whom were male, and 2590 were female. There were no significant differences in patient age ($p$=0.76) and sex ($p$=0.38) between the training and test datasets, nor was there a significant difference in site breakdown (all $p$>0.05) (Appendix p6).

Across the three deep learning architectures trained and evaluated, nnU-Net, SegResNet, and SwinUNETR, nnU-Net achieved the strongest fidelity. For patient-level lesion-detection metrics, nnU-Net achieved the highest balanced accuracy (0.830, 95% CI 0.791–0.872), exceeding both comparators on every metric, and the highest voxel-level enhancing-tumour Dice (paired Wilcoxon signed-rank test on per-case Dice: nnU-Net vs SegResNet, $p$<0.0001; nnU-Net vs SwinUNETR, $p$<0.0001) (Appendix p9-10). nnU-Net was therefore selected as the primary model for all subsequent analyses. Sample patients from all imaging cohorts are shown in Figure 3 below.

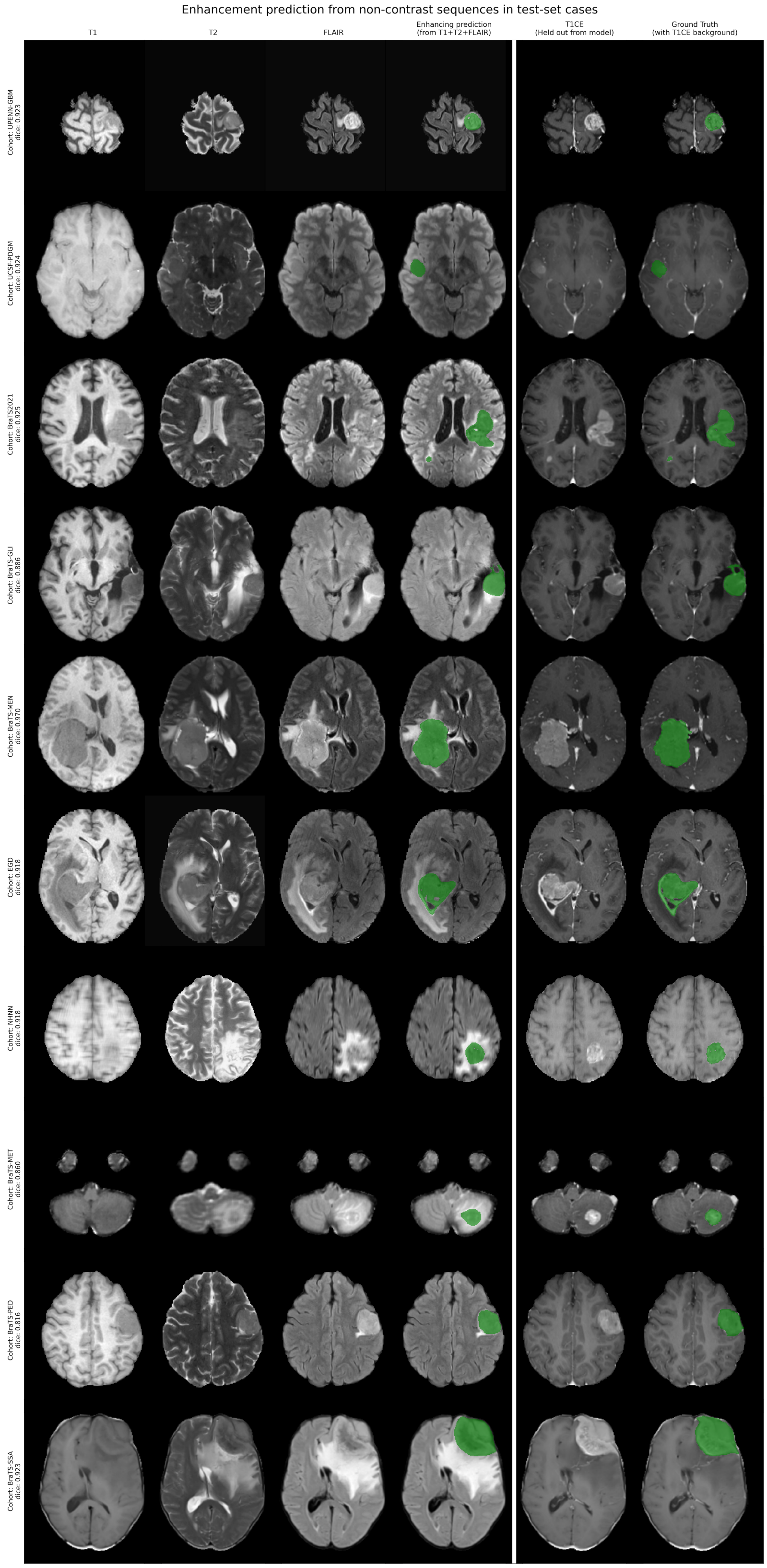

**Figure 3. Enhancement prediction from non-contrast sequences in test-set cases**. Representative test-set cases from each cohort demonstrating the model's performance. Each row shows a different case with six columns: T1w (T1-weighted), T2w (T2-weighted), T2/FLAIR (Fluid-Attenuated Inversion Recovery), model prediction of enhancing tumour from the unenhanced T1w, T2w and FLAIR images (green overlay on T2/FLAIR), T1CE (T1 contrast-enhanced, withheld during model training), and ground truth enhancing tumour segmentation (overlay on T1CE).

The radiologists, who were not specifically trained for this novel enhancement prediction task, achieved a mean balanced accuracy of 0.698 ± 0.072 across the 11 readers (bootstrap aggregate across all 1,100 blind reads; 95% CI, 0.671–0.725), which was lower than the model's performance. When these 1,100 reads were collapsed to a per-case majority vote on the 564 unique cases (Appendix p11), the radiologist panel's balanced accuracy rose to 0.717, an incremental benefit from multi-reader consensus. Of the 564 matched cases, the model and the radiologist agreed on 421 (both correct in 390, both incorrect in 31); of the 143 cases where they disagreed, the model was correct in 109 versus 34 for the radiologists (McNemar's exact two-sided test, $p<0.0001$). Overall, the model was correct in 499 of 564 cases (88.5% accuracy), compared with 424 (75.2%) for the radiologist majority; the corresponding balanced accuracies were 0.833 (95% CI 0.791–0.872) and 0.717 (0.670–0.763). Of the enhancing cases the radiologists did not correctly identify, the model correctly identified 79; conversely, of those the model did not correctly identify, radiologists identified 14. Agreement between readers on the multi-rater cases was fair (Cohen's $\kappa = 0.36$; observed agreement 68.3%), consistent with the difficulty of predicting enhancement from non-contrast imaging alone. Sample patients incorrectly reported by radiologists but correctly identified by the model are shown in Figure 4.

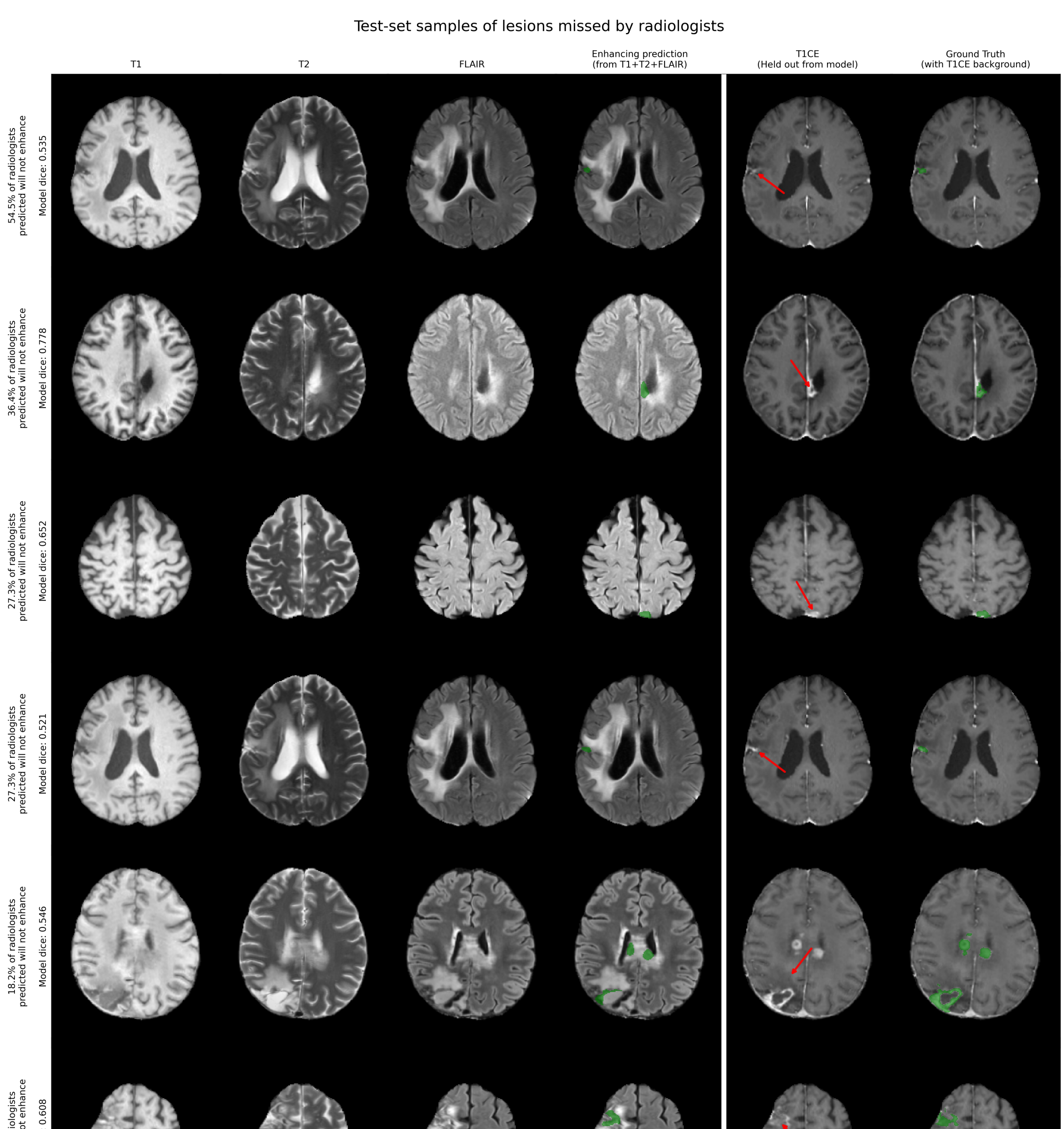

**Figure 4. Test-set samples of lesions missed by radiologists.** Representative test-set cases that radiologists wrongly predicted to contain no post-contrast enhancement when reviewing the non-contrast imaging. Each row shows a different case with six columns: T1w (T1-weighted), T2w (T2-weighted), FLAIR (Fluid-Attenuated Inversion Recovery), model prediction of enhancing tumour from the T1w, T2w and FLAIR (green overlay on FLAIR), T1CE (T1 contrast-enhanced, withheld during model training), and ground truth enhancing tumour segmentation (overlay on T1CE). Held-out T1CE images confirm that the model predictions align with actual contrast enhancement patterns.

Patient-level enhancement-detection performance, quantified by balanced accuracy (BA), was reasonably robust across the 1109 test set across all datasets, pathologies, countries, patient ages, and sexes, with BA ranging from 0.739 to 1.000 (Figure 5; Appendix p12-14). Slightly lower performances were observed in postoperative glioma resection cases (BA 0.825), paediatric presurgical tumours (BA 0.739, 95% CI 0.555–0.900), and within our internal NHNN cohort (BA 0.790, 0.717–0.856). Detection performance did not differ significantly between men (BA 0.816, 0.721–0.894) and women (BA 0.830, 0.738–0.909); performance stratified by patient age (n=565, recoverable for 50.9% of studies, as the adult BraTS cohorts are age-stripped at source) is shown in Figure 5.

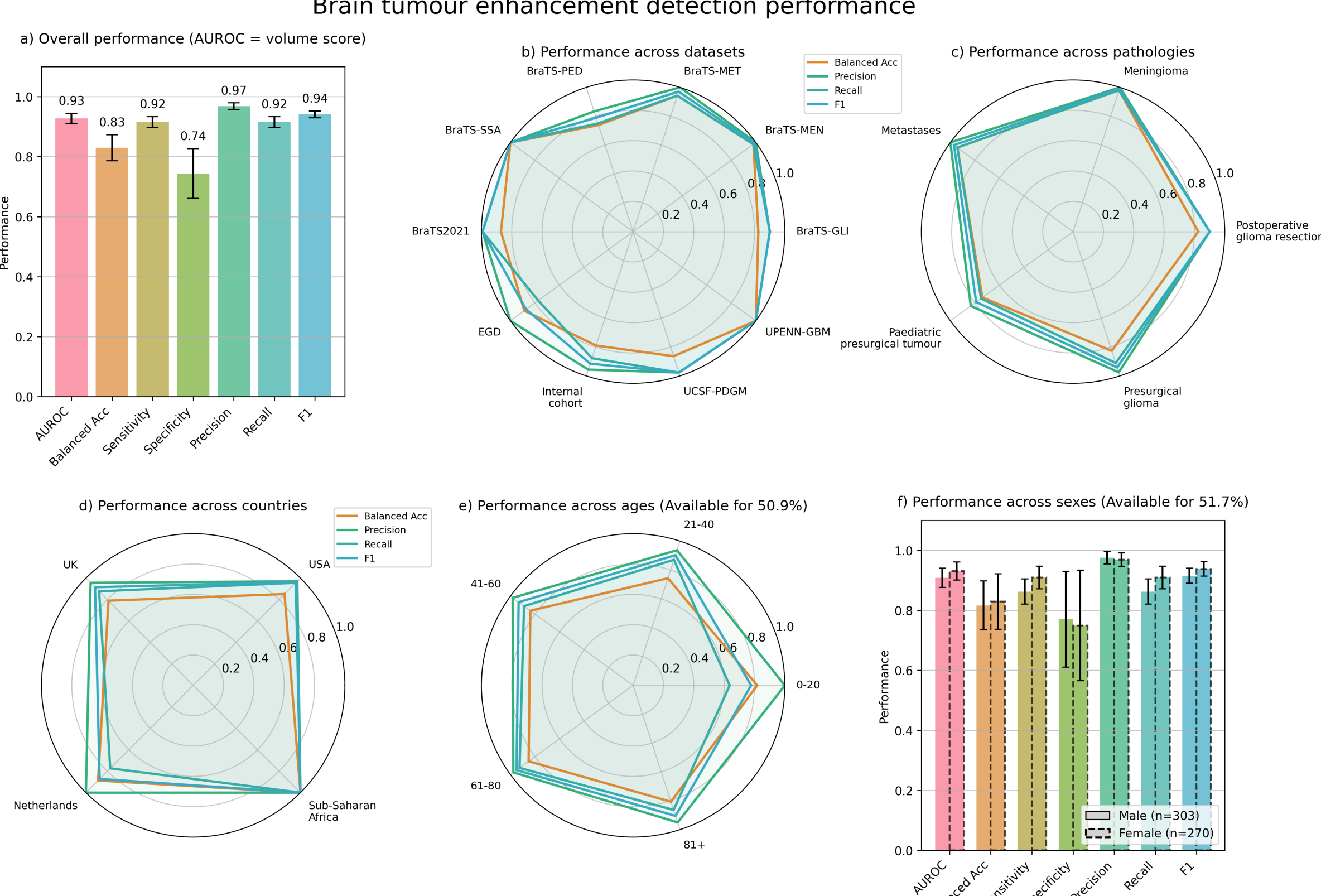

**Figure 5. Brain tumour enhancement detection performance.** Patient-level detection of enhancing tumour from non-contrast sequences stratified across the a) entire cohort, b) datasets, c) pathologies, d) countries, e) ages and f) sex. Bars in panels a and f are shown with bootstrapped 95% confidence intervals.

The voxel-level out-of-sample segmentation performance for nnU-Net, with mean ± SD Dice coefficient across the different components, was as follows: normal brain (0.987 ± 0.024), non-enhancing abnormal tissue (0.821 ± 0.233), and enhancing tumour (0.574 ± 0.319). Evaluating clinical performance thresholds among the 992 enhancing cases, we observed that 762 (76.8%) had a Dice score ≥ 0.3 (acceptable detection), 670 (67.5%) had ≥ 0.5 (good detection), and 498 (50.2%) had ≥ 0.7 (excellent detection).

For the 992 enhancement-positive cases, voxel-level Dice varied widely across cohorts — highest in research-grade glioma datasets (e.g., UPENN-GBM) and lowest in the older real-world NHNN cohort and the paediatric BraTS-PED cohort. Performance stratified by pathology and by country of origin is shown in Figure 5 and Appendix p13, and followed the same pattern, showing strongest fidelity for meningioma and weakest for paediatric tumours, with the UK (NHNN) 'real-world' dataset lower among countries.

The burden of enhancing tumour was accurately predicted by the model with an $R^2$ of 0.859 and a mean volumetric difference between the prediction and the ground truth of -1.30 $cm^3$ (Figure 6). There was a clear relationship between the case-level detectability of an enhancing lesion and its corresponding Dice score. Lesion detectability was more variable (approximately 50% success rate) for tiny lesions (~0.01 $cm^3$), but it correctly detected >90% of the time for lesions 2$cm^3$ or larger.

As commonly reported in segmentation tasks, voxel-level performance did show some variation with lesion volume (ANOVA *p*<0.0001) (Figure 6c), with Dice increasing monotonically from small to large lesions. There was also some variation in segmentation performance related to radiomic characterisation of lesions (ANOVA *p*<0.0001) (Figure 6d). A lesion's detectability was significantly related to its size, expressed by the logistic regression $y \sim 1.713\ \log_{10}(V) - 4.786$, where $V$ denotes lesion volume, *p*<0.0001, log-likelihood -361.464, Odds ratio per 10-times increase in volume 5.546 (95% confidence interval 4.389–7.009). There was no statistically significant relationship between model segmentation performance and the number of available samples per pathology.

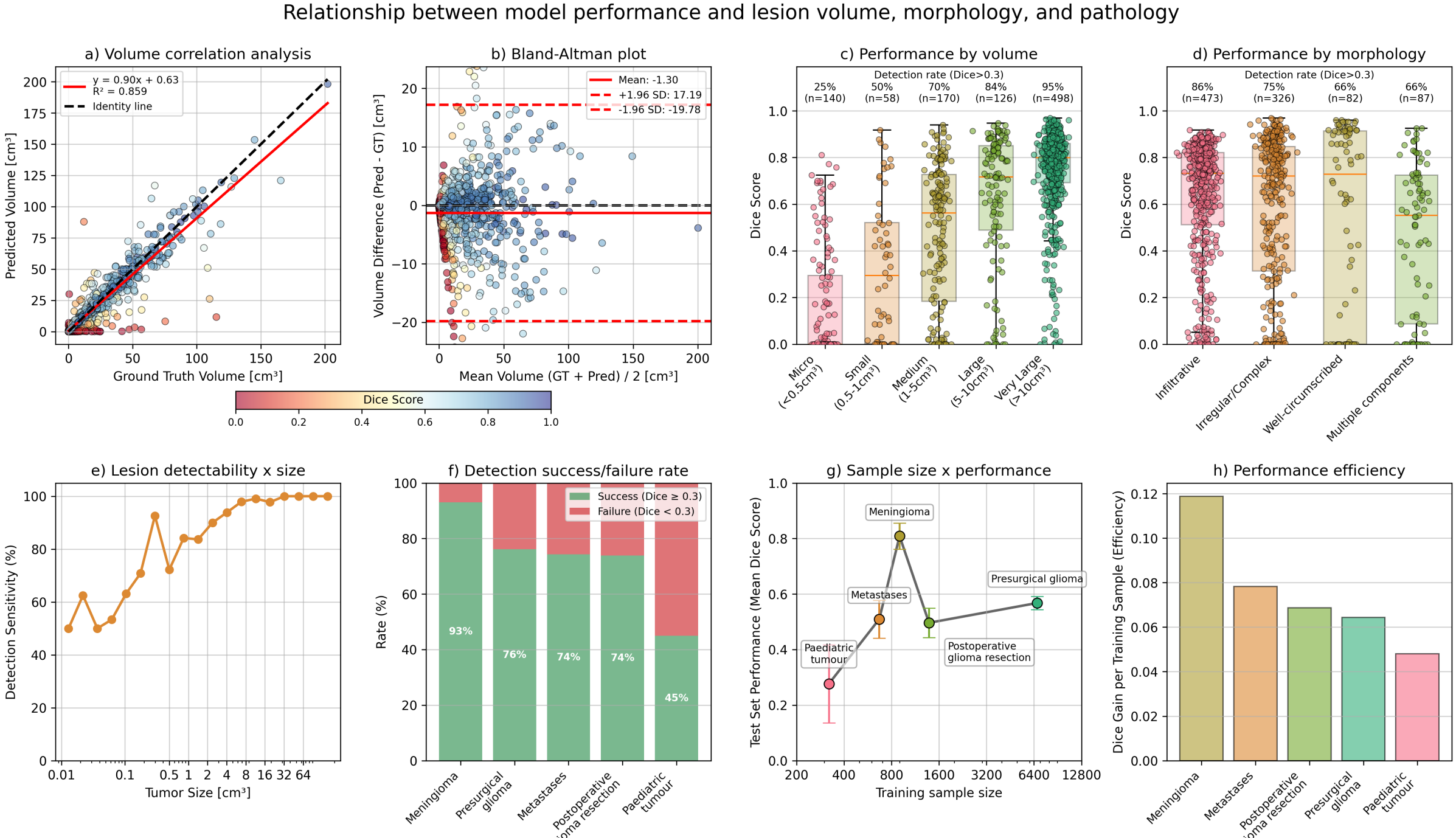

**Figure 6. Relationship between model performance and lesion volume, morphology, and pathology.** a) Volume correlation analysis showing strong agreement between predicted and ground truth enhancement (patient-level). b) Bland-Altman plot assessing agreement between predicted and ground truth volumes (patient-level), with mean difference (solid red line) and ±1.96 standard deviation limits (red dashed lines). c) Voxel-level segmentation performance (Dice coefficient) and successful detection rate (annotated percentage; success defined as voxel-level Dice ≥ 0.3) stratified by volume categories. d) Detection rate by radiomics-derived morphological patterns (patient-level binary detection). e) Patient-level lesion detection sensitivity as a function of enhancement volume (proportion of cases where any model-predicted enhancing voxels overlapped with ground truth). Note that panels c and e use different evaluation criteria (Dice ≥ 0.3 threshold vs binary overlap detection), which accounts for the apparent discrepancy in detection rates for small lesions. f) Detection success (green) and failure (red) rates by pathology type (patient-level). g) Relationship between training sample size and test set performance (Dice score) for each pathology type. h) Performance efficiency measured as Dice score gain per training sample, identifying pathologies with the highest learning efficiency.

Examination of cases with poor segmentation performance (Dice < 0.3) revealed several contributing factors that we systematically quantified. Of the cases failing to reach the acceptable detection threshold (Dice <0.3), the majority were attributable to small lesion volume (<0.5 cm³), followed by postoperative cases with complex enhancement patterns including surgical cavity enhancement, marked imaging artefact or suboptimal image quality in older acquisitions, and atypical enhancement patterns including faint or incomplete rim enhancement, or absent enhancement in lesions that would ordinarily be expected to enhance (the latter particularly noted in paediatric cases). A quantitative breakdown of failure categories is provided in Appendix p14. These observations suggest that model refinement should prioritise augmented training on challenging cases and potentially incorporate image quality assessment to flag unreliable predictions.

As an assessment of model uncertainty, we also evaluated model probability maps (i.e., prior to generating the binary segmentation map). Across the 1109 test set, the mean probability within the enhancing tumour prediction channel was 0.583 ± 0.301, with a mean entropy of 0.160 ± 0.089 (boundary fraction 0.075 ± 0.050). There were 0 cases with high uncertainty (i.e. with entropy > 0.5). The assessment of model uncertainty using probability maps is shown in Appendix p7. Maximal predicted probabilities were high across pathologies. Levene's test for equality of variances showed a significant difference (p=0.0021), demonstrating that model uncertainty was significantly more variable within the smaller pathology cohorts available. Probability variance did not differ significantly by sex or age category after multiple-comparison correction.

## Discussion

We developed and evaluated a deep learning model for detecting and segmenting enhancing tumour from non-contrast magnetic resonance imaging sequences alone. In the largest known study of its kind, using large-scale international data across ten different cohorts, five different disease categories, and four countries, our findings provide strong support for the ability of complex models to identify and delineate enhancing intracranial disease from non-contrast MRI alone. Although detecting contrast-enhancing disease is widely held to require contrast administration, our finding that a model can predict enhancement with >83% accuracy from only non-enhanced images suggests otherwise. Sufficiently capable AI models may leverage information from pre-contrast imaging that human experts cannot perceive. Such models may pave the way for reducing the use of contrast agents.

We envision the predominant real-world utility of this model as a triage or decision-support adjunct, rather than a complete replacement for contrast-enhanced imaging. Given the model's high sensitivity and the asymmetric cost of missing an enhancing lesion, the most defensible deployment is reciprocal: in workflows where only non-contrast sequences are protocolled, the model could flag patients in whom enhancement is likely so that a contrast-enhanced acquisition can be added, thereby avoiding a second hospital visit and minimising delays to patient care, while imaging predicted to be enhancement-free might, with appropriate caution, proceed without gadolinium — particularly in settings where repeated imaging is required (e.g., surveillance follow-up), where contraindications exist (renal impairment, allergy), or where minimising gadolinium exposure is prioritised (frequently repeated imaging). Any patient with an uncertain or positive prediction would proceed to standard contrast-enhanced imaging. Used in this direction, the tool could reduce gadolinium use while preserving the safeguard that suspected enhancement always triggers a confirmatory contrast study. The clinical value of this operating point is strongly prevalence-dependent. At the high enhancing-tumour prevalence of our enriched test set (89.4%; 992 of 1,109) the negative predictive value is limited (0.510), but rises to ~0.95 at the lower, surveillance-realistic prevalences (≈30%) at which a triage tool would most plausibly operate; the corresponding clinical-utility estimates (contrast studies spared and enhancing tumours missed per 1,000 scans) and full predictive values are given in Appendix p15-16. Predictive performance should therefore be interpreted against the prevalence in the intended deployment setting rather than in the development cohort.

Our model's performance, both as a patient-level lesion enhancement detector and a computer vision segmentation tool, is competitive with the broader literature. Previous internal research has demonstrated a Dice segmentation performance of 0.79 in delineating enhancing tumour in the more homogeneous Brats 2021 glioma cohort from T1w-, T2w-, and T2/FLAIR-weighted imaging alone[4]. However, this was evaluated in a substantially smaller and more uniform sampling space. Here, we report Dice segmentation performance for gliomas (cohort-dependent) of up to 0.824 on the UPENN-GBM dataset, but observe more variable performance depending on lesional and imaging heterogeneity and the quality of the available imaging data. Specifically, our deliberate inclusion of internal 'real-world' clinical data (NHNN cohort), which dates back to 2006, demonstrates substantial differences in imaging quality across MRI manufacturers, field strengths, and other associated parameters. Indeed, excluding this cohort raised rather than lowered overall performance (Appendix p17), indicating that its inclusion conservatively lowers our headline estimates. Other comparative studies include that of Kleesiek et al., who developed a Bayesian U-Net using T1w, T2w, FLAIR, diffusion-weighted imaging (b0, b1000, ADC), and susceptibility-weighted imaging (magnitude, phase, and minimum intensity projection) to produce virtual contrast maps in an 116-patient cohort[22], reporting strong discrimination for enhancing tumour detection in their validation set; our model achieved a balanced accuracy of 0.830 in an entirely held-out test set of 1,109 studies with many fewer imaging sequences provided. The inclusion of diffusion and susceptibility-weighted imaging by their group is particularly notable, not least because diffusion characteristics can assist radiologists in identifying tumour hypercellularity[23]. Regarding methods of reducing gadolinium dose (but not eliminating it entirely, as done here), Ammari et al. reported an 83% sensitivity for lesions larger than 10mm using deep learning for virtual contrast prediction, employing a one-quarter gadolinium dose[24], a task also undertaken by many others[5,25]. Murugesan et al. recently demonstrated similar feasibility in a single-institution glioma cohort[26]. Our study extends this approach to a substantially larger dataset (n=11,089) spanning multiple pathologies, age groups, and imaging protocols, thereby establishing broader generalisability. For direct comparison using consistent patient-level metrics, our model's 76.8% success rate (proportion of cases achieving Dice ≥ 0.3) should be interpreted in the context of our deliberately heterogeneous cohort. Most prior studies are typically conducted on smaller, more homogeneous, single-institution datasets with larger lesions[22,26], in which Dice values can be inflated[27]. Our cohort deliberately includes maximally heterogeneous real-world data spanning 10 datasets, 5 pathology types, 4 countries, and imaging from 2006–2024, including older, lower-quality acquisitions from the NHNN cohort.

This diversity, we would argue, while reducing aggregate success rates, provides a more realistic assessment of generalisability. The spread of voxel-level Dice across cohorts we interpret as the product of two distinct, partly separable factors: (1) distributional shift between sources — differences in scanner manufacturer, field strength, acquisition era, and annotation protocol, most pronounced in the older real-world NHNN data — and (2) genuine differences in lesion characteristics, since cohorts differ systematically in lesion size, morphology, and pathology, and smaller or more diffuse lesions are intrinsically harder to delineate. As in any machine-model study, lower aggregate performance in a given cohort reflects both how far that cohort is from the training distribution and how challenging its lesions are to segment.

An important consideration is the nature of our comparison with a human radiologist. The deep learning model was specifically trained on thousands of labelled examples to learn the mapping from non-contrast to contrast-enhancing appearances, whereas the radiologists were asked to perform a task outside their standard clinical training — without access to clinical history, prior imaging, or the referral indication. Radiologists are trained to interpret enhancement after gadolinium administration, not to predict it from intrinsic tissue characteristics alone. The radiologists' performance on this task is therefore not a reflection of their diagnostic capability in routine practice, where enhancement is read directly from post-contrast images; it indexes only the difficulty of a task deliberately designed to lie outside their standard clinical training. What the result does demonstrate, however, is that non-contrast sequences contain sufficient information to predict subsequent enhancement patterns in many cases, and that this ability was strongest with AI, a finding with implications for reducing gadolinium dependence and for understanding what tissue properties drive contrast uptake.

Equitable model performance across patient demographics and clinical subgroups is a necessary precondition for clinical deployment of any such tool[11,12,28]. Artificial intelligence is one of the fields experiencing the fastest growth in research, industry, and society, with many purported applications across medicine[29,30]. Yet, the equity assessment of software to ensure it benefits everyone is rarely quantified[11]. For these reasons, we assess the performance equity of our model as granularly as possible across the largest and most diverse available cohort. Our findings show that the model generally performs equitably across patient age and sex, although we found it less effective in paediatric populations. This weaker performance in paediatric cases is most likely partly due to our cohort containing substantially fewer paediatric cases than adults (only 352 of 11,089 [3.17%] paediatric, with the remainder adult). Although our sample is large in kind, paediatric patients were underrepresented compared to adults due to limited sample availability. Another potential, and likely combinatorial, factor is the inherent complexities of paediatric neuro-oncology, where enhancement patterns often vary significantly from those seen in adult lesions. Moreover, there is also considerable variability within the paediatric population itself, especially across the spectrum of paediatric low-grade gliomas and glioneuronal tumours. As a further strength, the inclusion of maximally heterogeneous data from as many sources as feasible will invariably maximise its generalizability, supported by other work highlighting the impact on model performance with distributional shift[31,32], and routes such as federated learning or widened data sharing to overcome it[32].

An alternative approach to predicting enhancement could employ quantitative parametric maps (e.g., T1/T2 relaxometry, perfusion CBV) instead of, or in addition to, qualitative weighted images. Such maps offer potential advantages: they are more standardised across scanners and directly measure tissue properties relevant to blood-brain barrier breakdown, which underlies contrast enhancement. However, quantitative parametric maps are often not routinely acquired in clinical practice, require longer acquisition times and additional post-processing, and are less available in older or resource-limited datasets. Notably, Kleesiek et al.[22] included diffusion-weighted and susceptibility-weighted imaging in their model and achieved strong discrimination in a much smaller cohort, whereas our model achieved a balanced accuracy of 0.830 using only standard clinical sequences (T1w, T2w, T2/FLAIR), suggesting that substantial enhancement-predictive information is already present in the conventional structural images. We would encourage future work that explores whether supplementing standard sequences with quantitative maps further improves performance, particularly for small or subtle lesions.

We deliberately framed this problem as segmentation rather than image-to-image synthesis or whole-image classification. A classifier returns only a study-level label and cannot localise disease, whereas synthesis of a full virtual post-contrast image is held to a far more demanding standard — every voxel of intensity and morphology must be faithfully reproduced — and current evidence indicates that generative synthesis is not yet achievable at the fidelity required at the clinical frontline. Morphology-preserving generative approaches[33] remain research tools rather than deployable diagnostics, and large foundation efforts such as NV-Generate-MR-Brain[34] similarly demonstrate plausible but not yet clinically validated synthesis. Segmentation occupies a useful middle ground for it simultaneously classifies (whether enhancing disease is present) and delineates (where it is), yielding an interpretable, spatially grounded output that is directly auditable against ground-truth enhancement without staking diagnostic safety on the fidelity of a synthesised image.

Our study has limitations. Firstly, the cohort disproportionately includes more adult than paediatric patients, and more larger than smaller lesions. Our research prioritised maximal inclusivity in all modelling tasks. To that end, the inclusion of *any* paediatric or small-lesion samples (even if markedly imbalanced relative to the availability of adult cohort data) was deemed preferable to excluding the population from this analysis. Future work should include larger cohorts with more diverse caseloads to improve model performance in these subpopulations, whether through new model training or fine-tuning using weights generated from adult data. To our strength, we tested this empirically within the analysis as a key part of equity assessment. We also included all data at our disposal, which encompassed global data from several countries and clinicodemographic populations, thereby providing greater generalizability than is often achieved in such studies. Secondly, though our model demonstrated strong sensitivity (0.915) for detecting an enhancing lesion in unenhanced images, its specificity was lower (0.744), with more false positives than false negatives. Future research could fine-tune these weights with additional data from further cohorts, including more non-enhancing tumours where available, given the imbalanced sample we had at our disposal. To our strength, one core purpose of such a model here would be to classify imaging sessions in which lesions are highly likely to enhance, so that post-gadolinium imaging could be forgone for the initial scan, follow-up imaging, or both. Therefore, a high sensitivity at the expense of specificity should arguably be preferred. A further limitation concerns the interpretation of the model's predicted probabilities. We report these as a descriptive characterisation of predictive entropy, not as calibrated probabilities. Formal calibration (for example, temperature scaling assessed with reliability diagrams) was not undertaken in this study. Post hoc calibration would be essential before any predicted probability is used to guide uncertainty-based deployment decisions, and is a priority for our further research. Thirdly, performance was weakest for small lesions, which accounted for the largest share of detection failures. This partly reflects internal class imbalance, for small enhancing targets are under-represented relative to large ones, but is also likely partly attributable to the known behaviour of Dice-loss[27], where overlap-based losses penalise small-structure errors disproportionately and bias models away from tiny targets. Future work should consider size-aware sampling, complementary detection-oriented losses, and lesion-size-stratified reporting.

In conclusion, we demonstrate that deep learning models can identify 'contrast-enhancing' intracranial disease in neuro-oncology when exposed to the non-contrast magnetic resonance imaging sequences alone. That a deep learning model can discern areas of pathological enhancement from non-enhanced images yields strong support for the advent of generative models in synthetically enhanced or low-dose enhanced post-contrast comparators[5-7]. In their current form, such approaches may be particularly valuable as potential decision-support or triage tools to minimise, rather than completely replace, the use of contrast-enhanced imaging. However, we hope future increases in data availability, algorithmic complexity, and high-performance computing may offer further opportunities to innovate in this space. These findings may also advance understanding as to the radiological basis for lesional enhancement, radiological pedagogy[35], and showcase the opportunities deep learning may bring for illuminating it as a means to reduce or even alleviate the need for contrast agents.

## Acknowledgements

JKR was supported by the Seed Grant funding programme of the European Society of Radiology (ESR) in collaboration with the European Institute for Biomedical Imaging Research (EIBIR) for undertaking this project. JKR is also supported by the Medical Research Council (MR/X00046X/1 & UKRI1389) and the British Society of Neuroradiology. HH and JKR are supported by the National Brain Appeal. PN is supported by the Wellcome Trust (213038/Z/18/Z). The UCLH NIHR Biomedical Research Centre supports JKR, HH, and PN. We thank the patients who contributed to the imaging in this study.

## Contributors

JKR and PN conceived the study and designed the methodology. JKR developed the software, conducted the formal analysis, and produced the figures and tables. JKR, SM, GP, and SB curated the imaging and metadata. JKR, SM, GP, SB, AB, AC, ID, DD, AH, HH, FJ, EL, DM, SO, and SW contributed to the radiologist reader study and case-level investigation. JKR and PN drafted the manuscript; all authors reviewed and approved the final version. JKR, HH, and PN were responsible for supervision, project administration, and funding acquisition. JKR and PN directly accessed and verified the underlying data, and JKR and PN were responsible for the decision to submit for publication.

## Data sharing

All imaging datasets used in this study, with the sole exception of the NHNN legacy cohort, are openly available under the terms of their original licences, and can be obtained from their respective repositories (BraTS-GLI, BraTS-MEN, BraTS-MET, BraTS-PED, BraTS-SSA, BraTS 2021, EGD, UCSF-PDGM, and UPENN-GBM)[13-21]. The NHNN cohort cannot be shared because the governing research ethics approval precludes onward dissemination of these clinical images. Model weights, the inference pipeline, and the code used for training, evaluation, and analysis are openly available at https://github.com/jamesruffle/enhancement_segmenter.

# Predicting brain tumour enhancement from non-contrast MR imaging with artificial intelligence: a multi-cohort retrospective diagnostic accuracy study - Appendix

## Methods

## Data

*Local data*

We included 3905 imaging sessions from 1364 unique patients who were imaged and treated for a glioma at our centre, the National Hospital for Neurology and Neurosurgery (NHNN), between 2006 and 2024, drawn from prospective histopathology record-keeping within the Institute's Department of Neuropathology. Having delineated a viable neuropathology cohort[1], we retrieved all magnetic resonance imaging (MRI) scans from the Picture Archiving and Communications System (PACS). We also recorded diagnostic and demographic data in these individuals.

*External data*

We acquired multiple open-source datasets to boost sample sizes and maximise the generalizability of results in neuro-oncology internationally. The datasets and sample sizes used were as follows: 1) The Brain Tumour Segmentation Challenge (BraTS) (n=1470)[2]; 2) the Erasmus Glioma Dataset (EGD) (n=774)[3]; 3) The University of Pennsylvania Glioblastoma (UPenn-GBM) cohort (n=665)[4]; 4) The University of California San Francisco Preoperative Diffuse Glioma (UCSF-PDGM) cohort (n=501)[5], 5) The Brain Tumour Segmentation – Post-treatment Challenge (n=1538)[6], 6) The Brain Tumour Segmentation – Metastases Challenge (n=739)[7], 7) The Brain Tumour Segmentation – Paediatric Challenge (n=352)[8], 8) The Brain Tumour Segmentation – Meningioma Challenge (n=1000)[9], and 9) The Brain Tumour Segmentation – Glioma in Sub-Saharan Africa Challenge (n=145)[10]. In instances of crossover between datasets (such as the inclusion of UCSF-PDGM[5]and UPenn-GBM[4], which both feature, in part, in BraTS2021[2]), we contacted the principal investigators of each study to clarify those specific cases and ensure that no participants were erroneously duplicated in our set. As a secondary quality control measure, any lesion whose appearance was highly correlated (voxel-wise r > 0.95) was flagged for cross-checking to ensure no duplicate participants. Paired demographics were variably available for these external repositories, the completeness of which was beyond our control.

## Image pre-processing

We acquired all available MRI sequences for all available imaging sessions from our local neuro-oncology cohort, with data ranging from 2006 to 2024. We deployed a locally developed MRI sequence classifier and artefact detector for initial image quality control and to curate imaging sequences[11]. We clamped MRI signal intensities at the 1$^{st}$ - 99$^{th}$ percentile to attenuate any spurious signal artefacts. All sequences were super-resolved to 1x1x1mm isotropic resolution using a locally developed generative super-resolution model[12]. Images were brain extracted with HD-BET[13], from which the brain extraction masks were also curated. The three non-contrast input sequences (T1-weighted, T2-weighted, and FLAIR-weighted images) were concatenated as separate channels along a fourth tensor dimension, providing each model with a four-dimensional multi-channel volumetric input.

We manually reviewed all raw data at all stages of the data pre-processing. Intensity- and scanner-harmonisation tools (for example, ComBat-style site/scanner normalisation) were considered but deliberately not applied. Because the central aim was generalisation across heterogeneous sites, scanners, and acquisition protocols, we elected to retain the inter-site and inter-scanner signal present in the multi-cohort data rather than regress it out, so that the model was trained and evaluated on the full distributional variability it is intended to encounter in deployment. Moreover, though harmonisation may well assist this task, harmonisation models can also introduce unpredictability in the context of pathology.

*Ground truth derivation*

Ground-truth enhancing-tumour masks were always generated from complete datasets — non-contrast T1-, T2-, and T2/FLAIR-weighted together with post-contrast T1-weighted imaging — which were available for all patients. For the external open-source datasets, the challenge-standard expert annotations supplied with each dataset were used directly; these were produced independently of our group. For the internal NHNN cohort, ground-truth labels were generated in-house with TumourSeg[14], our previously published, openly available tumour-segmentation model, whose architecture, weights, and training data are entirely separate from the three enhancement-prediction architectures evaluated here; every inferred NHNN label was then reviewed, and fine-tuned where required, by an expert neuroradiologist with seven years of neuro-oncology experience (JKR).

Whilst the NHNN labels were thus produced by a related in-house model and adjudicated by a single reviewer — which could in principle bias the NHNN voxel-level performance estimate — the patient-level presence or absence of pathological contrast enhancement was deemed plausibly objective and unbiased, since detecting it from post-contrast imaging is a standard radiological reporting skill. Critically, the 1,109 test cases were strictly held out from all training, including the training of the label-generation model, so that no test case contributed to either enhancement-prediction training or ground-truth-label generation.

Background (non-brain) labels were derived from brain extraction with HD-BET[13]. All sequences within an imaging session were acquired during the same scanning session, pre-processed using gold-standard approaches, and co-registered to MNI space (see *Image pre-processing*). Cases with substantial inter-sequence motion or registration failure on visual quality control were excluded. For postoperative cases, surgically related enhancement was distinguished from tumour enhancement by expert neuroradiological consensus within the source datasets.

*Nonlinear anatomical registration with enantiomorphic normalisation*

Having segmented lesions in the patient's native space, structural and lesion segmentation masks were nonlinearly registered to 1x1x1mm isotropic resolution MNI space with Statistical Parametric Mapping (SPM, v12) and enantiomorphic correction[15,16]. The advantage of enantiomorphic correction is that the risks of registration errors secondary to a lesion are minimised by leveraging a given patient's normal structural neuroanatomy on the unaffected contralesional hemisphere[15].

## Model development

Model architectures trained included nnU-Net (v2)[17], SegResNet[18], and SwinUNETR[19,20]. For nnU-Net, we trained the default 1000-epoch model and a 4000-epoch model with the large residual connection encoder modification, which represents a reasonable balance between computational cost and performance gain[21]. For SegResNet[18], architectural parameters included 16 output channels for the initial convolution layer, a dropout probability of 0.1, and ReLU activation. For SwinUNETR[19,20], architectural parameters included a feature size of 48, a dropout and attention dropout rate of 0.1. We used a learning rate of 0.001, the Adam optimiser, a cosine annealing learning scheduler, and Dice loss. For both SegResNet and SwinUNETR, we enabled a variety of augmentations, including foreground cropping and signal normalisation, random axis flipping, coarse dropout, intensity shift and scaling, bias field, histogram shifts, Gaussian smoothing, low resolution simulation, affine and 3D elastic deformations[22]. Data augmentations were enabled within nnU-Net per its creators' standard settings[17]. SegResNet and SwinUNETR models were trained for up to 4000 epochs, but with an early stopping patience of 100 epochs. Hyperparameters were tuned only within the training set, using internal validation of this set. For SegResNet and SwinUNETR, the dropout rate (0.1) and learning rate (0.001) reported above were selected on the basis of enhancing-tumour Dice achieved on the internal held-out validation split, rather than on any test-set performance. All three architectures were trained from random initialisation, with no pretrained weights used for any model. This eliminates the possibility of pretraining-derived data leakage, since no weights were derived from data overlapping the held-out test set. Model training was conducted on a Linux GPU-enabled workstation (NVIDIA RTX 6000 Ada Generation). Training times for a dataset of this size were substantial, at approximately 2000 hours (~83 days) per model. Total compute and training-hour budgets were matched across the three architectures. Within that fixed budget, nnU-Net's optimised data-loading and I/O permitted its default internal 5-fold cross-validation, whereas SegResNet and SwinUNETR were trained on a single held-out validation split due to compute demand constraints. We acknowledge that this does not constitute strict tuning-budget parity, since the architectures differed in how the matched budget was apportioned between cross-validation folds and training epochs.

## Evaluation

*Performance metrics*

We performed a randomised train-test split at the patient level, with 90% of patients allocated to training and 10% to model testing, ensuring no leakage between sets. The test set was locked until all model development was complete. For the NHNN cohort, where individual patients contributed multiple imaging sessions (up to 3-4 examinations per patient), the split was performed strictly at the patient level, ensuring that all examinations from a given patient were assigned to the same partition and maintaining complete patient-level independence between training and test sets. The unit of analysis was the imaging session rather than the patient, where a patient contributed more than one study (for example, separate preoperative and postoperative acquisitions), these represent distinct imaging presentations and were treated as independent observations. This partition was fixed for all model training runs to ensure maximal comparability. A leave-one-cohort-out (LOCO) training and evaluation scheme was considered but not adopted on the grounds of model training/inference equity. Excluding a cohort such as the Sub-Saharan Africa or paediatric datasets from training would have primed the model to underperform in precisely the populations in which equitable performance matters most, whereas excluding such a cohort from the test set would have removed the only empirical evidence of performance in those groups. We therefore retained all cohorts in both training and testing and instead offer cohort-stratified performance as a partial proxy for the cross-cohort generalisation that LOCO is intended to probe. This maximally inclusive, general-purpose design mirrors the trajectory of the BraTS challenge, which has moved from separate disease-specific tasks towards a unified 'Generalizability Across Tumours' objective; conversely, training separate models per cohort would require at least six distinct networks (preoperative glioma, postoperative glioma, Sub-Saharan low-field imaging, meningioma, metastases, and paediatric) and would forfeit the cross-cohort robustness that motivated pooling.

Metrics were formally defined as follows: Dice coefficient = $2|P \cap G| / (|P| + |G|)$; balanced accuracy = (sensitivity + specificity) / 2; sensitivity (recall) = TP / (TP + FN); specificity = TN / (TN + FP); precision = TP / (TP + FP); F1 = 2 × (precision × recall) / (precision + recall). For each metric, we specify whether it is computed at the patient level (binary classification of enhancement presence) or voxel level (spatial segmentation accuracy). Detection metrics (balanced accuracy, sensitivity, specificity, precision, and F1 score) were computed only for the enhancing class, as this was the primary clinical target. Balanced accuracy and macro-averaged metrics were used throughout because they are the standard, bias-resistant choice for evaluating class-imbalanced detection tasks[23,24]. The Dice coefficient was reported for all three foreground tissue classes (normal brain parenchyma, non-enhancing abnormal tissue, and enhancing tumour).

*Equity assessment*

Equity of model performance was assessed by disaggregating diagnostic performance across all available demographic and clinical subgroups — contributing dataset, site, country of acquisition, tumour diagnosis, patient age and sex — following the Fairboard framework[25]. Within each subgroup, we report the same patient-level detection metrics used throughout (balanced accuracy, sensitivity, specificity, precision, recall and F1, together with the area under the receiver-operating-characteristic curve), so that performance can be compared directly across groups. Subgroups containing only enhancement-positive patients have undefined specificity, in which case balanced accuracy reduces to sensitivity. No subgroup-specific probability recalibration was performed. We also quantified the model detection rate in relation to radiomic features of enhancement, determining the relationship between segmentation performance and whether it was infiltrative, irregular/complex, well-circumscribed, solitary or multiple. The radiomic categories were defined based on quantitative shape features extracted from 3D binary masks of enhancing brain lesions. The following features were calculated: 1) Number of connected components from 3D connected component analysis; 2) Sphericity, calculated as $\frac{\pi^{\frac{1}{3}}(6V)^{\frac{2}{3}}}{A}$, where $V$ denotes volume and $A$ is surface area; 3) Solidity: calculated as the ratio of actual lesion volume to convex hull volume; 4) Compactness, calculated as $\frac{A^3}{(36\pi V^2)}$, where $A$ is surface area and $V$ is volume; 5) Surface-to-Volume Ratio, the surface area divided by volume; and 6) Elongation, ratio of largest to smallest eigenvalues from principal component analysis of lesion voxel coordinates. Based on these quantitative features, lesions were classified into the following radiomic categories: 1) Multiple – imaging sessions with ≥3 connected components, or two components where the largest component represents <80% of total enhancement volume; 2) Well-circumscribed

single - a single dominant lesion with sphericity > 0.7 and solidity > 0.9; 3) Infiltrative single - a single dominant lesion with sphericity < 0.5 or solidity < 0.7; 4) Irregular/complex single – a single dominant lesion with intermediate features of sphericity 0.5 - 0.7 and solidity 0.7 - 0.9.

## Software use

The following Python software and models were used for model development and downstream analyses: Fairboard[25], Matplotlib[26], MONAI[27], Nibabel[28], Nilearn[29], NumPy[30], pandas[31], PyTorch[32], seaborn[33], scikit-learn[34], and TumourSeg[14].

# Results

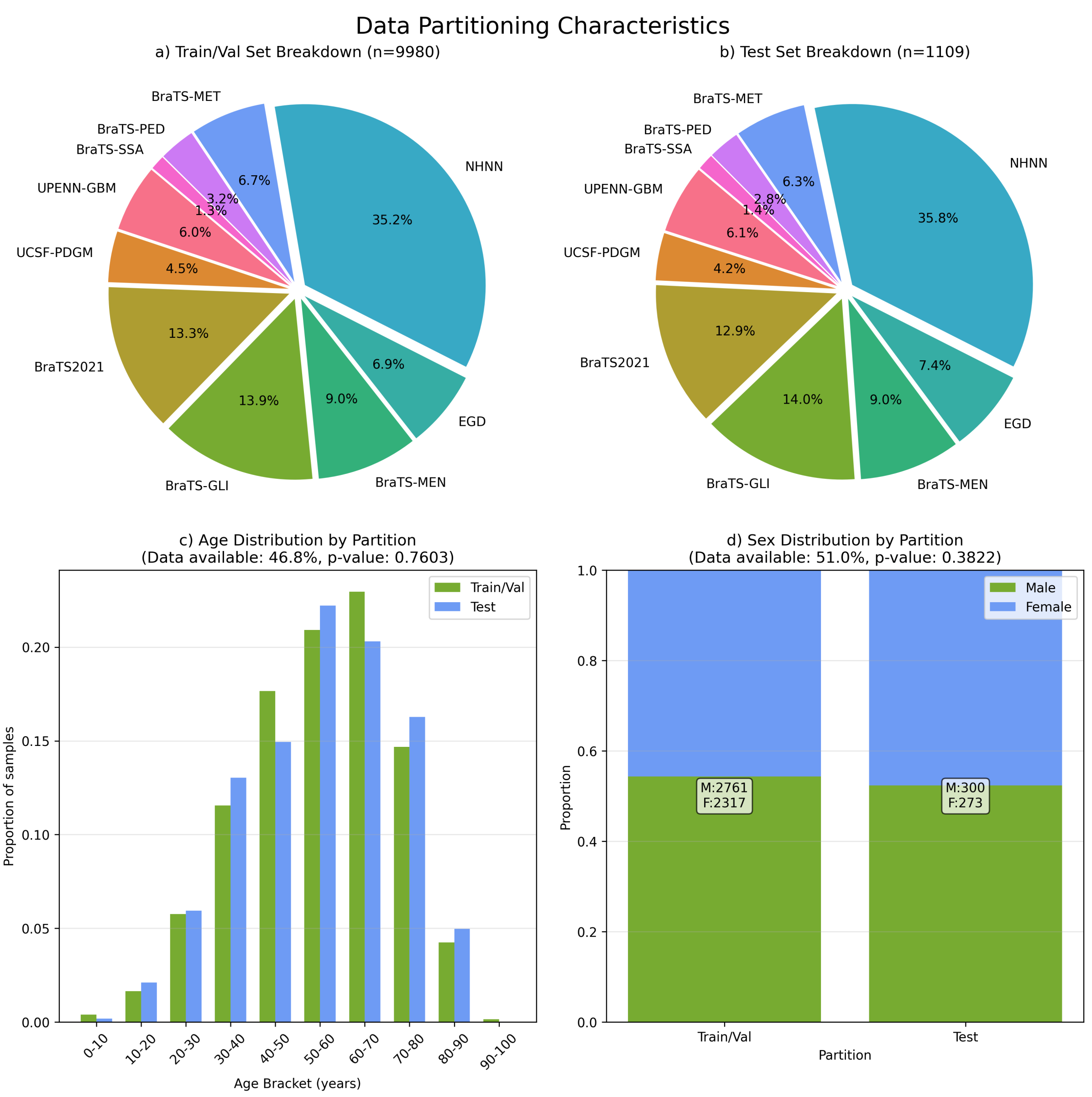

**Supplementary Figure 1. Data partitioning characteristics**. a) Training and validation set composition (n=9,980) showing proportional representation across all cohorts. b) Test set composition (n=1,109), maintaining equivalent proportions to training data, ensuring unbiased evaluation. c) Age distribution comparison between training/validation and test set partitions. d) Sex distribution comparison between training/validation and test set partitions.

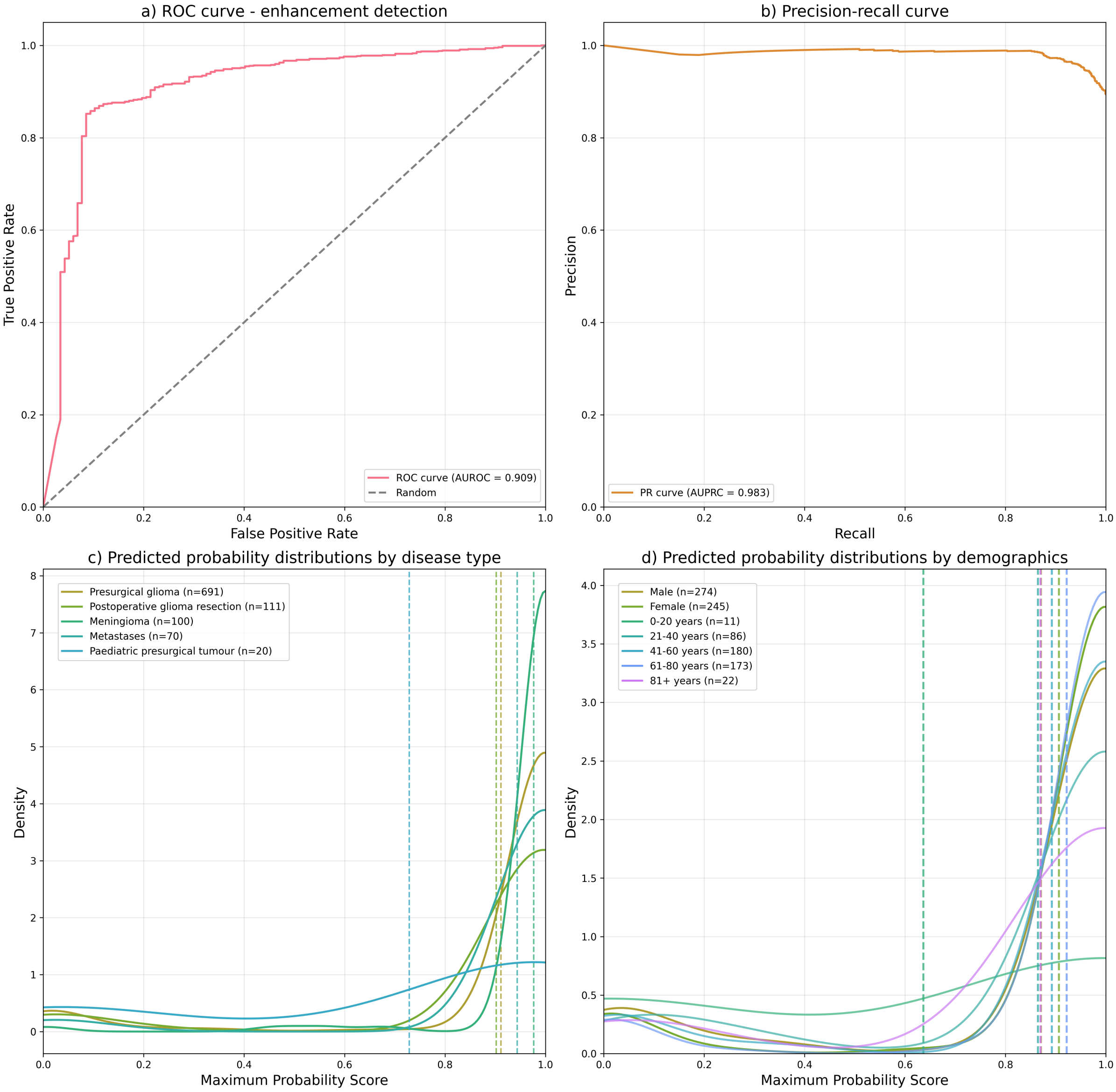

**Supplementary Figure 2. Model discriminative performance analysis**. a) Receiver Operating Characteristic (ROC) and b) Precision-Recall curve for enhancement detection. c-d) Predicted probability distributions stratified by c) pathology type and d) demographics.

Patient-level detection differences were confirmed by paired case-level bootstrap: nnU-Net exceeded SegResNet by a balanced-accuracy difference of +0.081 (95% CI 0.026–0.137, $p$=0.0034) and SwinUNETR by +0.119 (0.063–0.174, $p$<0.0001) (Supplementary Table 2). This advantage was driven by sensitivity (nnU-Net vs SegResNet Δsensitivity +0.154, $p$<0.0001; vs SwinUNETR +0.169, $p$<0.0001) and F1 (+0.092 and +0.105 respectively, both $p$<0.0001), whereas specificity differences were not significant (vs SegResNet +0.009, $p$=0.94; vs SwinUNETR +0.068, $p$=0.24). The patient confusion matrices for enhancement detection between the model and the radiologist are also provided below in Supplementary Table 4.

| Cohort | Country | Pathology | Imaging sessions (train/test) | Unique patients | Age available, n (%) | Age, mean (SD), y | Sex (M/F/Unknown) |
|---|---|---|---|---|---|---|---|
| UPENN-GBM | USA | Glioma | 665 (597/68) | 665 | 665 (100%) | 62.4 (12.4) | 402/263/0 |
| UCSF-PDGM | USA | Glioma | 501 (454/47) | 501 | 501 (100%) | 56.9 (15.0) | 299/202/0 |
| BraTS 2021 | USA | Glioma | 1470 (1327/143) | 1470 | 0 | N/A | 0/0/1470 |
| BraTS-GLI | USA | Glioma (post-op) | 1538 (1383/155) | 1538 | 0 | N/A | 0/0/1538 |
| BraTS-MEN | USA | Meningioma | 1000 (900/100) | 1000 | 0 | N/A | 0/0/1000 |
| EGD | Netherlands | Glioma | 774 (692/82) | 774 | 725 (93.7%) | 56.7 (14.6) | 492/281/1 |
| NHNN | UK | Glioma | 3905 (3508/397) | 1364 | 3296 (84.4%) | 52.8 (17.2) | 1868/1844/193 |
| BraTS-MET | USA | Metastases | 739 (669/70) | 739 | 0 | N/A | 0/0/739 |
| BraTS-PED | USA | Paediatric | 352 (321/31) | 352 | 0 | N/A | 0/0/352 |
| BraTS-SSA | Nigeria | Glioma | 145 (129/16) | 145 | 0 | N/A | 0/0/145 |
| Overall | — | Mixed | 11,089 (9,980/1,109) | 8,548 | 5,187 (46.8%) | 55.0 (16.4) | 3,061/2,590/5,438 |

**Supplementary Table 1. Cohort demographics and baseline characteristics**. Age is in years, mean (SD), among sessions with recoverable age ('Age available'); the public BraTS challenge cohorts are age- and sex-stripped at source. Sex is shown as male/female/unknown.

| Metric | nnU-Net | SegResNet | SwinUNETR | Expert radiologists (n=11) |
|---|---|---|---|---|
| Balanced accuracy | **0.830 ± 0.021** | 0.748 ± 0.021 | 0.711 ± 0.023 | 0.698 ± 0.072 |
| Sensitivity (recall) | **0.915 ± 0.009** | 0.762 ± 0.013 | 0.746 ± 0.013 | 0.749 ± 0.076 |
| Specificity | **0.744 ± 0.040** | 0.735 ± 0.040 | 0.676 ± 0.043 | 0.647 ± 0.151 |
| Precision | **0.968 ± 0.006** | 0.961 ± 0.007 | 0.951 ± 0.008 | 0.680 ± 0.091 |
| F1 score | **0.941 ± 0.005** | 0.849 ± 0.009 | 0.836 ± 0.009 | 0.713 ± 0.056 |

**Supplementary Table 2. Patient-level enhancement detection performance on the held-out test set of 1,109 studies.** Model values are bootstrap mean ± SD (1,000 resamples, seed 42). Radiologist values are the aggregate across 1,100 blind reads (11 readers × 100 cases each; SD across per-rater metrics).

| Metric | nnU-Net | SegResNet | SwinUNETR |
|---|---|---|---|
| Dice coefficient ↑ | **0.574 ± 0.319** | 0.523 ± 0.288 | 0.421 ± 0.277 |
| Jaccard index (IoU) ↑ | **0.467 ± 0.290** | 0.402 ± 0.251 | 0.306 ± 0.226 |
| Sensitivity (recall) ↑ | 0.581 ± 0.337 | 0.627 ± 0.286 | **0.635 ± 0.297** |
| Precision ↑ | **0.709 ± 0.259** | 0.516 ± 0.308 | 0.389 ± 0.291 |
| Hausdorff distance, mm ↓ | **11.00 [6.56–22.60]** | 16.43 [9.62–41.45] | 30.71 [13.93–89.60] |
| 95% Hausdorff distance, mm ↓ | **4.01 [2.24–9.91]** | 6.16 [3.61–16.85] | 11.57 [5.74–31.88] |
| Average symmetric surface distance, mm ↓ | **1.25 [0.71–2.83]** | 2.03 [1.22–5.23] | 3.97 [1.97–10.21] |
| Mean surface distance, mm ↓ | **1.18 [0.69–2.47]** | 1.95 [1.19–4.53] | 3.66 [1.92–8.10] |

**Supplementary Table 3. Voxel-wise segmentation performance for the predicted enhancing-tumour class, by architecture.** Voxel-wise segmentation performance for the predicted enhancing-tumour class on the 1,109-study held-out test set, comparing the three candidate architectures. Overlap and voxel-classification metrics (Dice, Jaccard/IoU, sensitivity, precision) are mean ± SD over all test samples with ground-truth enhancement. Distance metrics (Hausdorff, 95% Hausdorff, average symmetric surface distance, mean surface distance) are medians [IQRs], in mm, across studies in which all three models produced a non-empty enhancement prediction. Arrows indicate the favourable direction and bold marks the best value per row. Overlap (Dice/IoU) and voxel counts were obtained using the nnU-Net evaluation routine; surface-distance metrics were obtained using SimpleITK (exact Maurer distance transform). nnU-Net was superior to both SegResNet and SwinUNETR on Dice, IoU, precision and all four distance metrics (paired Wilcoxon signed-rank, all $p<0.0001$); voxel sensitivity did not differ significantly between the networks (p=0.46 and p=0.13); the comparators nonetheless tend to over-segment, evidenced by their substantially lower precision and larger surface distances.

**A. nnU-Net model**

| GT ↓ / Prediction → | Model: enhancement | Model: no enhancement |
|---|---|---|
| GT: enhancement (n=447) | 412 | 35 |
| GT: no enhancement (n=117) | 30 | 87 |

**B. Expert radiologists (per-case majority vote)**

| GT ↓ / Prediction → | Radiologists: enhancement | Radiologists: no enhancement |
|---|---|---|
| GT: enhancement (n=447) | 347 | 100 |
| GT: no enhancement (n=117) | 40 | 77 |

**Supplementary Table 4. Patient-level confusion matrices for binary enhancement detection.** Test computed on the 564 unique test cases reviewed by the expert-radiologist panel (447 GT-positive, 117 GT-negative). (A) nnU-Net model. (B) Radiologist majority vote (blind reads, no AI assistance; 359 single-rater, 205 multi-rater cases). **N.B.** (i) Because each radiologist independently sampled 50 enhancement-positive and 50 enhancement-negative cases, the 564 unique cases reviewed contain a higher proportion of ground-truth-negative cases than the full 1,109-study test set (20.7% vs 10.6% GT-negative). This enrichment accounts for the small difference between the model's balanced accuracy here (0.833) and in Supplementary Table 2 (0.830), where the full test-set prevalence applies. (ii) Of the 564 cases, 359 were reviewed by a single radiologist and 205 by two or more. Among the 205 multi-reviewed cases, raters were unanimous in 114 (55.6%) and disagreed in 91 (44.4%). Pairwise inter-rater agreement was fair (Cohen's $\kappa = 0.36$; observed agreement 68.3%), consistent with the difficulty of predicting enhancement from non-contrast imaging alone. (iii) 23 cases had an even number of raters split exactly 50/50, which in the main analysis were resolved as 'positive' (enhancement predicted) to emulate maximal clinical sensitivity.

| Row | n | BA (95% CI) |
| --- | --- | --- |
| **Model architecture** | | |
| nnU-Net | 1,109 | 0.830 (0.791–0.872) |
| SegResNet | 1,109 | 0.748 (0.705–0.786) |
| SwinUNETR | 1,109 | 0.711 (0.666–0.756) |
| | | |
| **Subgroup (*nnU-Net only*)** | | |
| Paediatric (BraTS-PED) | 31 | 0.739 (0.555–0.900) |
| NHNN | 397 | 0.790 (0.717–0.856) |
| Under 30 (exact age) | 85 | 0.765 (0.577–0.906) |
| 30 and over (exact age) | 480 | 0.833 (0.764–0.893) |
| Male | 303 | 0.816 (0.721–0.894) |
| Female | 270 | 0.830 (0.738–0.909) |

**Supplementary Table 5. Balanced accuracy with 95% CI by architecture and subgroup.** Patient-level balanced accuracy with percentile bootstrap 95% CIs (seed 42, 1,000 resamples). Architecture rows compare the three candidate networks, whereas the subgroup rows are computed for the deployed nnU-Net model.

| Cohort | N | Dice (mean ± SD) | Mean Dice (95% CI, bootstrap) | Successful detection rate (Dice ≥ 0.3) | Successful detection rate (Dice ≥ 0.5) | Successful detection rate (Dice ≥ 0.7) |
|---|---|---|---|---|---|---|
| UPENN-GBM | 68 | 0.824 ± 0.085 | 0.804–0.843 | 100.0% | 98.5% | 94.1% |
| BraTS-MEN | 100 | 0.809 ± 0.239 | 0.756–0.852 | 93.0% | 91.0% | 85.0% |
| BraTS2021 | 139 | 0.766 ± 0.185 | 0.735–0.795 | 96.4% | 91.4% | 83.5% |
| BraTS-SSA | 16 | 0.717 ± 0.151 | 0.642–0.781 | 100.0% | 87.5% | 81.2% |
| UCSF-PDGM | 43 | 0.704 ± 0.304 | 0.609–0.790 | 83.7% | 83.7% | 76.7% |
| EGD | 75 | 0.533 ± 0.334 | 0.452–0.605 | 73.3% | 68.0% | 50.7% |
| BraTS-MET | 70 | 0.510 ± 0.291 | 0.441–0.578 | 74.3% | 64.3% | 30.0% |
| BraTS-GLI | 111 | 0.496 ± 0.283 | 0.444–0.553 | 73.9% | 61.3% | 29.7% |
| NHNN | 350 | 0.423 ± 0.313 | 0.391–0.456 | 62.0% | 46.9% | 26.3% |
| BraTS-PED | 20 | 0.277 ± 0.321 | 0.138–0.418 | 45.0% | 35.0% | 15.0% |
| **Total (weighted)** | **992** | **0.574 ± 0.319** | 0.554–0.594 | **76.8%** | **67.5%** | **50.2%** |

**Supplementary Table 6. Voxel-level segmentation performance (enhancing-tumour Dice) on the held-out test set.** Table stratified by source cohort (n = 992 ground-truth-positive cases; 117 GT-negative cases excluded as Dice is undefined without ground-truth enhancement). Bootstrap 95% CIs on the mean.

| **Cohort (test-set N)** | **False positive** | **Small lesion (<0.5 cm$^3$)** | **Paediatric** | **Postoperative** | **Other/atypical** | **Total** | **Rate (%)** |
|---|---|---|---|---|---|---|---|
| **BraTS-PED (n=31)** | 3 | 3 | 8 | 0 | 0 | 14 | 45.2 |
| **NHNN (n=397)** | 14 | 62 | 0 | 0 | 71 | 147 | 37.0 |
| **BraTS-GLI (n=155)** | 11 | 12 | 0 | 17 | 0 | 40 | 25.8 |
| **BraTS-MET (n=70)** | 0 | 9 | 0 | 0 | 9 | 18 | 25.7 |
| **EGD (n=82)** | 0 | 11 | 0 | 0 | 9 | 20 | 24.4 |
| **UCSF-PDGM (n=47)** | 1 | 5 | 0 | 0 | 2 | 8 | 17.0 |
| **BraTS-MEN (n=100)** | 0 | 3 | 0 | 0 | 4 | 7 | 7.0 |
| **BraTS2021 (n=143)** | 1 | 0 | 0 | 0 | 5 | 6 | 4.2 |
| **UPENN-GBM (n=68)** | 0 | 0 | 0 | 0 | 0 | 0 | 0.0 |
| **BraTS-SSA (n=16)** | 0 | 0 | 0 | 0 | 0 | 0 | 0.0 |
| **Total (N=1109)** | 30 | 105 | 8 | 17 | 100 | 260 | 23.4 |

**Supplementary Table 7. Failure category breakdown by cohort on the held-out test set.** Counts of test-set cases from the 1,109-study held-out set whose enhancement detection and voxel-level Dice for the enhancing-tumour class fell below 0.3 (n = 260 of 1,109 failing cases; cohort denominators are shown in the column headers).

| Prevalence | Sensitivity | Specificity | PPV | NPV | Contrast administrations avoided per 1,000 | Enhancing tumours missed per 1,000 |
|---|---|---|---|---|---|---|
| 89.4% (test) | 0.915 | 0.744 | 0.968 | 0.510 | 155 | 76 |
| 50% | 0.915 | 0.744 | 0.781 | 0.898 | 414 | 42 |
| 30% | 0.915 | 0.744 | 0.605 | 0.953 | 546 | 25 |

**Supplementary Table 8. Clinical utility at the deployed operating point across prevalences.** Estimated at the fixed deployed operating point (sensitivity (0.915) and specificity (0.744) are properties of the model and therefore constant across prevalences; positive and negative predictive values derived by Bayes' theorem and clinical-impact counts expressed per 1,000 scans. The 89.4% row reflects the held-out test set prevalence, with 50% and 30% represent lower-prevalence (more surveillance-realistic) settings.

| Prevalence | PPV | NPV |
|---|---|---|
| 10% | 0.284 | 0.988 |
| 20% | 0.472 | 0.972 |
| 50% | 0.781 | 0.898 |
| 80% | 0.935 | 0.687 |
| 89.4% (test) | 0.968 | 0.510 |

**Supplementary Table 9. Negative and positive predictive value as a function of disease prevalence.** Derived by Bayes' theorem at the fixed deployed operating point (sensitivity 0.915, specificity 0.744). NPV is limited at the high test-set prevalence (89.4%) but rises sharply at lower, surveillance-realistic prevalences, whereas PPV shows the converse pattern.

| Set | n (GT+) | BA (95% CI) | Sensitivity | Specificity | Precision | F1 | Voxel Dice (mean ± SD) |
|---|---|---|---|---|---|---|---|
| Full test set (N=1,109) | 1,109 (992) | 0.830 (0.791–0.872) | 0.915 | 0.744 | 0.968 | 0.941 | 0.574 ± 0.319 |
| NHNN-excluded external (n=712) | 712 (642) | 0.854 (0.798–0.900) | 0.936 | 0.771 | 0.974 | 0.955 | 0.656 ± 0.291 |

**Supplementary Table 10. Patient-level performance with and without the internal NHNN cohort.** Patient-level nnU-Net metrics on the full held-out test set and on the external-only subset after excluding all internal NHNN cases. Balanced accuracy is higher on the external-only data than on the full set, bounding any concern about model-assisted-label circularity in the model's favour. Bootstrap 95% CIs (seed 42, 1,000 resamples); voxel Dice computed on ground-truth-positive cases only.